\documentstyle[12pt]{article}
\input epsf
\begin{document}

\newpage
\bigskip
\hskip 3.7in\vbox{\baselineskip12pt
\hbox{NST-ITP-99-}
\hbox{hep-th/}}

\bigskip\bigskip

\centerline{\large \bf S-Matrices from AdS Spacetime}

\bigskip\bigskip

\centerline{\bf
Joseph Polchinski\footnote{joep@itp.ucsb.edu}} 
\medskip
\centerline{Institute for Theoretical Physics}
\centerline{University of California}
\centerline{Santa Barbara, CA\ \ 93106-4030}

\bigskip

\begin{abstract}
\baselineskip=16pt
In the large-$N$ limit of $d=4$, $N=4$ gauge theory, the dual AdS space
becomes flat.  We identify a gauge theory correlator whose large-$N$ limit is
the flat space S-matrix.
\end{abstract}
\newpage
\baselineskip=18pt
\setcounter{footnote}{0}
The Maldacena dualities~\cite{malda,corr} relate string theory in various
near-horizon geometries to gauge and other quantum field theories.  If
correct, these give nonperturbative definitions of string
theory in these backgrounds.  For example, one could {\it in
principle} simulate the quantum field theory on a large enough computer,
which is the criterion originally set forth by Wilson~\cite{wilson} for a
nonperturbative definition in the case of quantum field theory.  Further, in
the large-$N$ limit of each field theory the curvature and field strengths of
the dual geometry vanish, and so it should be possible to extract any
property of the flat spacetime string theory.  In this note we pursue this
idea, addressing the following question: what quantity in large-$N$ gauge
theory corresponds to the flat spacetime string S-matrix?

The description of the flat-spacetime limit is highly nonuniversal and
noncovariant: by taking different quantum field duals, different
kinematics, and different processes one obtains very different
constructions of the S-matrix.  We present here only one such
construction, from a limit of $AdS_5 \times S^5$.  It is to be hoped that
some universal and covariant description can be extracted from the
large-$N$ limit, and the kinematics of the present paper may be a
useful step in this direction.  Other discussions of holomorphy and flat
spacetime appear in~\cite{otros}.

There have been many recent discussions of scattering processes in AdS
spacetime~\cite{op1,op2,many}.  The present work certainly overlaps these,
but we do not know of work that directly addresses the question discussed
here.  Incidentally, there appear to be a belief (also nonuniversal) that
the S-matrix cannot be extracted from anti-de Sitter space even in a limit;
we do not understand these arguments, but the explicit construction here may
help to clarify the issues.

We consider $D=4$ $N=4$ $SU(N)$ gauge theory on $S^3$ with Minkowski time,
which is dual to IIB string theory on the whole of $AdS_5 \times
S^5$~\cite{horoog}.  One could also consider the space $R^3$, but the dual
geometry is incomplete and one would have to arrange the kinematics
carefully to avoid losing outgoing particles.  The metric of $AdS_5$ is
\begin{eqnarray}
ds^2 &=& R^2 \Biggl[ - (1+r^2) dt^2 + \frac{dr^2}{1+r^2} + r^2 d\Omega_3^2
\Biggr] \nonumber\\
&=& R^2 \Biggl[ - (1+r^2) dt^2 + d{\bf x} \cdot d{\bf x} -
\frac{({\bf x} \cdot d{\bf x})^2}{1+r^2}
\Biggr]  \ ,
\end{eqnarray}
where $R^4 = 4\pi \alpha'^2 g_{\rm s} N$ and $r^2 = {\bf x} \cdot {\bf x}$.
Points on the boundary $S^3$ will be labelled by unit four-vectors ${\bf e}$.
Because the metric contains a factor of $R^2$, distances as measured by an
inertial observer differ by a factor of $R$ from coordinate distances; we
will always refer to the former as `proper,' and similarly for momenta.
We will always dimensionally reduce on the $S^5$ factor,
producing an effective mass term in $AdS_5$.

Consider first the case of particles that are massless in $AdS_5$.  The
geodesic motion
\begin{equation}
{\bf x} = {\bf e}\tan t \ ,\quad
p_t = \omega\ ,\quad {\bf p} =  \frac{\omega {\bf e}}{1+r^2}\ ,
\label{traject}
\end{equation}
begins at the boundary point $- {\bf e}$ at $t = -\pi/2$, reaches
the origin at
$t=0$, and returns to the boundary point $+ {\bf e}$ at $t =
+\pi/2$.  A particle on a second geodesic, reflected by $ {\bf e} \to
-{\bf e}$, will intersect the first at the origin.  A scattering process
would have a proper center-of-mass energy-squared
\begin{equation}
s = -g^{tt} (2p_t)^2 = 4\omega^2/R^2 \ .  \label{mands}
\end{equation}
If the particles scatter into $n$ massless outgoing particles, the latter
will still reach the boundary at $t = +\pi/2$ but at general points of the
asymptotic $S^3$.  Thus it is possible to probe the general massless
scattering with sources on the boundary at $t = -\pi/2$ and detectors on
the boundary at $t = +\pi/2$.  By holding the external proper momenta of
the process fixed as $R \to \infty$, one obtains the flat spacetime
scattering amplitude.  This corresponds to holding the detector angles
fixed while, by eq.~(\ref{mands}), 
$\omega \propto R$.  In terms of the underlying string parameters, we wish
to obtain the 10-dimensional theory with given $g_{\rm s}$.  This
corresponds to $N \to \infty$ with $g_{\rm s}$ fixed.  In summary,
\begin{equation}
N \to \infty\ ,\quad\mbox{$g_{\rm s}$, $\alpha'$,
and
$s$ fixed\ ,}\quad R = (4\pi \alpha'^2 g_{\rm s} N)^{1/4}\ ,
\quad \omega = \frac{1}{2}Rs^{1/2} \ . \label{limit}
\end{equation}

We now use this reasoning to give an LSZ-like prescription for the
massless particle S-matrix.  We will assume initially that the particles
propagate freely except for their interaction near the origin, and then discuss
corrections. There is one annoying complication. Thus far we have described the
scattering classically, specifying both the positions and momenta of the
external particles.  For ordinary flat spacetime scattering one can put the
external particles in momentum eigenstates, because the scattering location is
irrelevant.  In the present case, in order to obtain definite kinematics in
the flat limit, the scattering must occur in a known position due to the
position-dependence of the metric.  Thus we must resort to wavepackets.

Since $\omega$ is large in the limit of interest a WKB approximation can be
used.  For simplicity we consider scalar particles.  The details are given in
the appendix; we present here the results.  There is a solution
$\phi_{\omega{\bf e}}$ to the free wave equation, which follows the classical
trajectory~(\ref{traject}) with an uncertainty $\omega^{-1/2}$ in $\bf x$.
To be precise, in the neighborhood of the origin
\begin{equation}
\phi_{\omega{\bf e}}(t,{\bf x}) \approx F_{\omega{\bf e}}(t,{\bf x})
e^{-i\omega(t - {\bf
 e} \cdot {\bf x})}\ , \label{packeta}
\end{equation}
with $F_{\omega{\bf e}}(t,{\bf x})$ a smooth envelope of width $\omega^{-1/2}$
centered on the trajectory ${\bf x} = {\bf e} t$.  The coordinate
width $\omega^{-1/2}$ goes to zero at large $N$, so the width of the
packet is small compared to the AdS radius, while the proper width
$R\omega$ goes to infinity.  The overlap region of the packets is
well-localized compared to the AdS scale, as desired, while the uncertainty in
the proper momentum is $R^{-1} \omega^{-1/2}$ and goes to zero.  Thus the
scattering of the packets approaches the flat space process.  At
$r \to \infty$,
\begin{equation}
\phi_{\omega {\bf e}}(t,{\bf x}) \approx \Bigl[ G_{-} (t + \pi/2,|{\bf \hat x +
e}|) + G_{+} (t - \pi/2,|{\bf \hat x - e}|) \Bigr] e^{-i\omega t}
\ .\label{packetb}
\end{equation} 
The two terms represent the beginning and ending of the trajectory on the
boundary.  The functions $G_{\pm}$ are of width $\omega^{-1/2}$ in both time
and angle, and as discussed in the appendix are related in a simple way to 
$F_{\omega {\bf e}} (t,{\bf x})$.  Incidentally, the solution
$\phi_{\omega {\bf e}}$ has no reflected piece at $t > \pi/2$ or $t < -
\pi/2$.

Consider now the current
\begin{equation}
j_{{\omega {\bf e}} \mu} = \hat\phi \partial_\mu \phi_{\omega {\bf e}}
- \phi_{{\omega {\bf e}}} \partial_\mu \hat \phi
\ .
\end{equation}
Here the hat denotes the field {\it operator} in the effective bulk quantum
field theory.  Continuing to ignore interactions away from the origin, this
current is conserved.  Define then
\begin{equation}
\alpha_{\omega {\bf e}} = \int_S dA\, n^\mu j_{{\omega {\bf e}} \mu}\ .
\label{mode}
\end{equation}
Due to current conservation, we can take for an incoming particle either
of the surfaces $S_1$ and $S_2$ shown in figure~1, each of which intersects the
packet before the scattering region.
\begin{figure}
\begin{center}
\leavevmode
\epsfbox{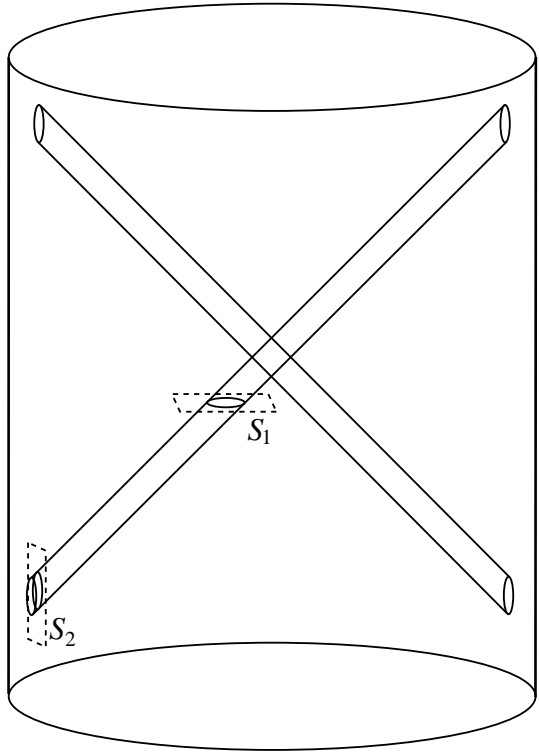}
\end{center}
\caption[]{Surfaces $S_1$ and $S_2$ intersecting an incoming wavepacket.}
\end{figure}
For an outgoing particle there are
corresponding surfaces after the scattering.  The spacelike surface $S_1$
is taken to lie in the flat region near the origin as $N \to \infty$.  The
integral~(\ref{mode}) then defines a flat-space creation operator for an
incoming particle; with the envelope function~$F_{\omega {\bf e}}$ omitted
this would give a covariantly normalized plane wave.  For the timelike region
$S_2$ at the boundary we can use the dictionary~\cite{op1,op2}
\begin{equation}
\lim_{r \to\infty} r^4 \hat\phi(t,{\bf x}) = \hat {\cal O}(t,{\bf \hat x})\ ,
\label{field}
\end{equation}
where $\hat {\cal O}$ is the corresponding operator\footnote{We continue to
ignore interactions, but in fact we believe that the relation~(\ref{field})
will hold at least perturbatively in the interacting theory \cite{benpol}.} in
the gauge theory on $S^3$.  We define $\hat\phi$ to have canonical
normalization as a five-dimensional field, so that the relation~(\ref{field})
defines an implicit normalization for $\hat {\cal O}$.

The integral on $S_2$ can therefore be expressed as an operator in the gauge
theory.  The analysis can immediately be extended to particles with masses of
order the Kaluza--Klein scale $R^{-1}$ (which are therefore massless in ten
dimensions).\footnote{
S-matrices for particles with nonzero ten-dimensional masses cannot be
studied in this way; note, however, that the IIB string theory has no BPS
particle states.}
 Classically these do not reach the boundary but do come
very close, reaching $r \sim
\omega$.  They can still be created by $\hat\phi$ at the boundary but
with an appropriate tunneling factor.  In fact, the boundary
behavior~(\ref{packetb}) acquires a term
\begin{equation}
r^{\nu - 2}\ ,\quad \nu = (M^2 R^2 + 4)^{1/2}\ , \label{nu}
\end{equation}
while the factor of $r^4$ in the operator relation~(\ref{field}) becomes $r^{2
+ \nu}$.  The integral on $S_2$, and the corresponding integral for outgoing
particles, are then
\begin{eqnarray}
\alpha_{{\omega {\bf e}}-} &=& 2\nu \int dt\,d^3{\bf \hat x}\,
G_{-} (t + \pi/2,|{\bf \hat x + e}|) e^{-i\omega t} \hat {\cal O}(t,{\bf \hat
x}) \nonumber\\
\alpha_{{\omega {\bf e}}+} &=& 2\nu \int dt\,d^3{\bf \hat x}\,
G_{+} (t - \pi/2,|{\bf \hat x - e}|) e^{-i\omega t} \hat {\cal O}(t,{\bf \hat
x}) \ .
\label{boundmode}
\end{eqnarray}
These express the bulk creation and annihilation operators in terms of the
operators in the boundary gauge theory.  Incidentally, the solution
$\phi_{\omega {\bf e}}$ is nonnormalizable, while the wave operator
$\hat\phi$ couples to the normalizable modes that appear in the quantization
of the field in AdS space.  For any mass the product of these behaves as
$r^{-4}$ at large $r$; combined with $r^{-1}$ from $\partial_r$, $r$
from $n^r$, and 
$r^4$ from the metric, the integrals defining $\alpha_{{\omega {\bf e}}\pm}$
are $r$-independent.

The flat-space S-matrix is then
\begin{equation}
S(I,O) = \lim_{N \to\infty} \Phi^{-1} \Biggl\langle \prod_{i \in I} 
\alpha_{{\omega_i{\bf e}_i}-} \ \prod_{j \in O} 
\alpha_{{\omega_j{\bf e}_j}+} \Biggr\rangle\ . \label{lszp}
\end{equation}
Here $I$ and $O$ denote the sets of incoming and outgoing particles,
The proper energy of each particle is $\omega / R$ and the proper momentum is
$\omega {\bf e}/ R$. The
expectation value is in the gauge theory on $S^3$, with the
operators~(\ref{boundmode}). The factor $\Phi$ accounts for the overlap of
wavepackets,
\begin{equation}
\Phi = \int dt\,d^4{\bf x}\,\int_{S^5} d^5x'\, \prod_{i \in I \cup O}
F_{\omega_i {\bf e}_i}(t,{\bf x}) \psi_i(x')  \ .
\end{equation}
Here $\psi_i(x')$ is the normalized wavefunction on $S^5$; for an $SO(6)$
singlet the net contribution of the compact space is $(V_{S^5})^{(2-n)/2}$.
The proper momenta in the $S^5$ direction are of order $1/R$ and so vanish in
the limit: the scattering process is restricted to a five-dimensional plane.
To study processes with nonzero momenta in the $S^5$ directions would require
the use of high representations of $SO(6)$, scaling with $N$.

Consider now corrections to free propagation, as depicted in figure~2.
\begin{figure}
\begin{center}
\leavevmode
\epsfbox{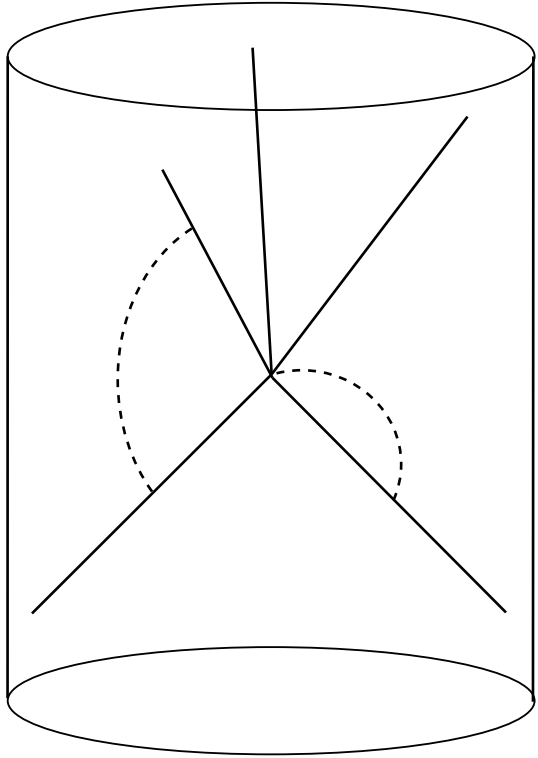}
\end{center}
\caption[]{Interactions (dashed lines) correcting the free propagation of
wavepackets.}
\end{figure}
Of course, these same processes are present in flat space, so they will change
the S-matrix formula only to the extent that interactions at distances of
order the horizon size are important.  The characteristic size of the
scattering region is set by the external momenta and so scales as
$\omega^{-1}$.  Thus, the interaction corrections will only be dangerous if
they are IR divergent.  In five dimensions even with massless
particles amplitudes at generic momenta are IR convergent.\footnote{In
lower dimensions one should in any case be considering not the S-matrix but an
appropriate IR-finite inclusive amplitude.}  This is a flat-spacetime
result but in general one expects that IR diverenges are exacerbated by
positive curvature and reduced by negative curvature~\cite{calwil}.  This
appears to be the case here as well, for example from examination of the gauge
boson propagator in ref.~\cite{dhofre}.

Ref.~\cite{op2} discusses various more subtle ways in which the
relation~(\ref{field}) may fail due to interactions.  However, we believe
that our result for the S-matrix is robust.  A local perturbation in the gauge
theory is expected to correspond to a local disturbance on the boundary of AdS
space; this is fully consistent with interactions in the Euclidean
case~\cite{corr}, for example.  This disturbance will propagate into the
interior of AdS space as a particle or multiparticle state; appropriate
kinematics then produces the S-matrix.  We have used local
fields~(\ref{field}) to derive the LSZ formula and of course the theory in the
bulk is not a local field theory.  However, we have used the field relation
only in a very weak sense, essentially its vacuum to one-particle matrix
element --- there is no assumption that field theory, or locality,
holds in the interaction region.  In particular, we see no obstacle to
assuming that the LSZ expression holds for arbitrarily large proper energies,
of order the string scale, the Planck scale, or beyond.
Note however that the proper energy is held
fixed as $N \to \infty$.

It would be interesting to subject the S-matrix result~(\ref{lszp}) to
various tests.  However, many of its required properties, such as $SO(9,1)$
invariance, will not be manifest but instead must be taken as predictions for
the behavior of the gauge theory.  It may be possible to analyze the pole
structure using the OPE in the gauge theory.

The final expression~(\ref{lszp}) in the gauge theory involves three
energy--momen\-tum scales: order 1 (in coordinate units) from the separation of
the sources and the curvature of $S^3$, order $\omega \sim N^{1/4}$ from the
incoming and outgoing waves, and order $\omega^{1/2}$ from the envelope
function.  One could also include a simpler object, in which the envelope
functions are omitted and so one integrates the sources and detectors over
times and angles (perhaps with spherical harmonics).  This still gives the
flat spacetime S-matrix but now with some
average over external momenta because the kinematics of the scattering depends
on its location, and also with the possible complication of multiple
scattering from the periodicity of motion in AdS spacetime.  Thus, 
flat-spacetime physics is obtained in the large-$N$ limit of a two-scale
object, with momenta of order~1 and of order
$g_{\rm s} N^{1/4}$.  Further, the large momenta appear only in the time
direction of the gauge theory.  Incidentally, there seems to be no
simple distance--energy relation such as there is in one-scale
processes~\cite{malda,holo}.

\section*{Appendix} 
\setcounter{equation}{0}
\def\theequation{A.\arabic{equation}}
To analyze wavepackets with $\hat e = (1,0,0,0)$ it is convenient to use 
coordinates $(\rho,{\bf y})$ with $\bf y$ a three-vector, as defined by
\begin{equation}
{\bf x} = (\tan \rho, {\bf y}/\cos\rho)\ .
\end{equation}
The classical trajectory of interest is simply
\begin{equation}
 \rho = t\ ,\quad {\bf y} = 0\ .
\end{equation}
In these coordinates the metric is 
\begin{equation}
ds^2 = \frac{R^2}{\cos^2\rho}
\Biggl[ -(1+y^2) dt^2 + d\rho^2 + d{\bf y} \cdot d{\bf y} - \frac{ ({\bf y}
\cdot d{\bf y})^2}{1 + y^2} \Biggr]\ .
\end{equation}
The d'Alembertian is
\begin{equation}
R^2 \nabla^2 = -\frac{\cos^2\rho}{1+y^2} \partial_t^2 + \cos^5\rho
\partial_\rho (\cos^{-3} \rho \partial_\rho) + \cos^2\rho ( \partial_{\bf y}
\cdot
\partial_{\bf y} + \partial_{\bf y} \cdot {\bf y}\,{\bf y} \cdot \partial_{\bf
y} )
\ .
\end{equation}

We seek a solution of the form
\begin{equation}
\phi = \exp\Bigl[ i\omega f(t,\rho) - \omega y^2 g_1(\rho) - \omega
(t - \rho)^2 g_2(\rho) + h(\rho) \Bigr]\ .
\end{equation}
We are interested in the case $\omega \gg 1$ and so make an analysis of
geometric optics (WKB) type.  The first term in the exponent is the rapidly
varying phase.  The second and third terms produce the envelope in space and
time; it follows that $\bf y$ and $t - \rho$ are of order $\omega^{-1/2}$.

Expanding $\nabla^2 \phi = 0$ in powers of $\omega$, the term of order
$\omega^2$ is
\begin{equation}
\omega^2 [ (\partial_t f)^2 - (\partial_\rho f)^2 ] = 0
\end{equation}
with solution $f = \rho - t$.  At order $\omega$,
\begin{equation}
\omega \cos^2 \rho \Bigl[ -\omega y^2 - 2i \omega y^2 g'_1 - 2i
\omega (t-\rho)^2 g'_2 + 2i h' + 3i \tan\rho - 2g_1 + 4\omega y^2
g_1^2 \Bigr] = 0\ ,
\end{equation}
where the prime is a $\rho$ derivative.  Thus,
\begin{eqnarray}
g_1' &=& \frac{i}{2} - 2i g_1^2\ , \nonumber\\
g_2' &=& 0\ , \nonumber\\
h' &=& -\frac{3}{2} \tan\rho - i g_1\ . 
\end{eqnarray}
These are readily integrated.  A simple particular solution, which we will
use henceforth, is
\begin{equation}
g_1 = g_2 = \frac{1}{2}\ ,\quad h = \frac{3}{2} \ln\cos\rho - \frac{i\rho}{2}\
.
\end{equation}
At the origin this solution is of the form~(\ref{packetb}) with 
\begin{equation}
F_{\omega {\bf e}}(t,{\bf x}) = \exp\biggl\{ - \frac{\omega}{2} \Bigl[
x_\perp^2  + (t - {\bf e} \cdot {\bf x})^2 \Bigr] \biggr\}\ .
\end{equation}
Here ${\bf x}_\perp$ is the part of $\bf x$ that is orthogonal to $\bf e$.

Very near the boundary, $r \sim \omega$, the WKB approximation breaks down.
Here we can match onto the large-$r$ behavior
\begin{equation}
\phi = A(t,{\bf\hat x}) \frac{e^{-i\omega t}}{r^2} H^{2,1}_2(\omega/r)\ ,
\label{bessel}
\end{equation}
where the variation of $A(t,{\bf\hat x})$ is slow compared to the remaining
factors.  The superscripts $1,2$ on the Bessel function refers to the behavior
at $t = \pm \pi/2$.  In the regime $\omega \gg r \gg 1$ both the large-$r$
and WKB expressions are valid and so we can match, with the result
\begin{equation}
A(t,{\bf\hat x}) = - i e^{\pm i\pi\omega/2} (\pi\omega/2)^{1/2}
\exp\biggl\{ - \frac{\omega}{2} \Bigl[
|{\bf\hat x} \mp {\bf e}|^2  + (t \mp \pi/2)^2 \Bigr] \biggr\}\ .
\end{equation}
The $r \gg \omega$ behavior of the Bessel function then gives the wavepacket
on the boundary,
\begin{equation}
G_{\pm} (\tau,\theta) = - e^{\pm i\pi\omega/2} (2/\omega)^{3/2}
\pi^{-1/2}
\exp\biggl\{ - \frac{\omega}{2} [
\theta^2  + \tau^2 ] \biggr\}\ .
\end{equation}

For scalars with masses of order the Kaluza--Klein scale $R^{-1}$, the
trajectory and WKB analysis are unaffected for $r$ less than $\omega$.  The
effect of the mass is then simply to change the order of the Bessel function
to $\nu$, and the result for the wavepacket is
\begin{equation}
G_{\pm} (\tau,\theta) =  e^{\pm i\pi (\omega+\nu) /2}
(2/\omega)^{\nu - 1/2}
\Gamma(\nu) \pi^{-1/2} \exp\biggl\{ - \frac{\omega}{2} [
\tau^2  + \theta^2 ] \biggr\}\ .
\end{equation}

\section*{Acknowledgments} 

I would like to thank V. Balasubramanian, T. Banks, I. Bena, S. Giddings, D.
Gross, G. Horowitz, N. Itzhaki, P. Pouliot, and M. Srednicki for discussions.
This work was supported in part by NSF grants PHY94-07194 and
PHY97-22022.

\end{document}